\pdfoutput=1

\documentclass[12pt]{iopart}

\usepackage{amscd,amssymb,a4}
\usepackage{graphicx,xspace}

\usepackage[left=3cm,top=3cm,right=3cm,bottom=3cm]{geometry}
\usepackage{color}

\begin{document}

\newcommand{\red}{\color{red}}
\newcommand{\blue}{\color{blue}}
\newcommand{\todo}[1]{\textbf{To do: #1}}
\newcommand{\be}{\begin{equation}}
\newcommand{\ee}{\end{equation}}
\newcommand{\beq}{\begin{equation}}
\newcommand{\eeq}{\end{equation}}
\newcommand{\bea}{\begin{eqnarray}}
\newcommand{\eea}{\end{eqnarray}}
\newcommand{\rar}{\rightarrow}
\newcommand{\lar}{\leftarrow}
\newcommand{\ra}{\right\rangle}
\newcommand{\la}{\left\langle }
\renewcommand{\d}{{\rm d }}
\newcommand{\m}{{\tilde m }}
\newcommand{\p}{\partial}
\newcommand{\nn}{\nonumber }

%
%
\bibliographystyle{KAY}

\newcommand{\atanh}
{\operatorname{atanh}}

\newcommand{\ArcTan}
{\operatorname{ArcTan}}

\newcommand{\ArcCoth}
{\operatorname{ArcCoth}}

\newcommand{\Erf}
{\operatorname{Erf}}

\newcommand{\Erfi}
{\operatorname{Erfi}}

\newcommand{\Ei}
{\operatorname{Ei}}

\newcommand{\sgn}{{\mathrm{sgn}}}

\def\be{\begin{equation}}
\def\ee{\end{equation}}

\def\bea{\begin{eqnarray}}
\def\eea{\end{eqnarray}}

\def\e{\epsilon}
\def\l{\lambda}
\def\d{\delta}
\def\o{\omega}
\def\cb{\bar{c}}
\def\Li{{\rm Li}}

\title[Quench in Lieb-Liniger model and KPZ equation]{Interaction quench in a Lieb-Liniger model and the KPZ equation with flat initial conditions}

\author{Pasquale Calabrese}

\address{Dipartimento di Fisica dell Universit\`a di Pisa and INFN, 56127 Pisa Italy}

\author{Pierre Le Doussal}

\address{CNRS-Laboratoire de Physique Th\'eorique de l'Ecole Normale Sup\'erieure\\
24 rue Lhomond, 75231 Paris Cedex-France
}

\date{\today}

\begin{abstract}
Recent exact solutions of the 1D Kardar-Parisi-Zhang equation make use of the 1D integrable 
Lieb-Liniger model of interacting bosons. For flat initial conditions, it requires the knowledge of
the overlap between the uniform state and arbitrary exact Bethe eigenstates. The same 
quantity is also central in the study of the quantum quench from a 1D non-interacting Bose-Einstein 
condensate upon turning interactions. We compare recent advances in both domains,
i.e. our previous exact solution, and a new conjecture by De Nardis et al.. This leads to new
exact results and conjectures for both the quantum quench and the KPZ problem. 
\end{abstract}

\maketitle

\section{Introduction}

The Lieb Liniger model for $n$ bosons in one dimension with a delta function two body interaction 
is based on one of the simplest integrable quantum many-body hamiltonian 
\be
H_n = -\sum_{\alpha=1}^n \frac{\partial^2}{\partial {x_\alpha^2}}  + 2 c \sum_{1 \leq  \alpha < \beta \leq n} \delta(x_\alpha - x_\beta).
\label{LL}
\ee
This model was solved a long time ago using the Bethe Ansatz \cite{ll}, the eigenstates and eigen-energies being known.
Since then, its properties have been much studied, although the 
difficulties in manipulating the Bethe eigenstates often make these studies technically challenging. 
These properties are quite different in the attractive case and the repulsive case. 
In the repulsive case there is a proper fixed-density thermodynamic limit \cite{ll}, and the bosons 
form a 1D superfluid with quasi long-range order, a sea with particle-hole excitations, 
and collective modes \cite{ll2} which, at low energies, are well described by the bosonisation theory \cite{caz}
(i.e. equivalently by Luttinger liquid theory). 
In the attractive case, the ground state is a bound state of all $n$ bosons, and the excitations
are obtained by splitting it into a collection of quasi-independent, smaller bound states, which behave {\it almost} 
as free particles \cite{m-65}. 

Recent experimental realisations \cite{exp} with dilute cold atomic gases revived the interest in this 
well-known integrable model. 
Furthermore, even from the theoretical perspective, new difficult questions emerged from 
two apparently unrelated contexts, namely 
(i) the Kardar-Parisi-Zhang (KPZ) equation for the noisy growth of an 1D interface \cite{KPZ} and 
(ii) quantum quench problems in 1D cold atoms.
The relation between these two apparently very different subjects is provided by the Cole-Hopf transformation which maps
the KPZ equation into the directed polymer problem: the height field of the KPZ interface at time $t$ can be written as
$h \propto \ln Z$, $Z$ being the partition sum of directed paths of length $t$ in a random potential which encodes the external noise feeding into the KPZ equation.
It was noted a while ago \cite{kardareplica,bb-00} that the integer moments $\overline{Z^n}$ can 
be represented as matrix elements of the (imaginary time) evolution operator $e^{- t H_n}$ of the Lieb-Liniger model
in the attractive regime. 
It then appears that the KPZ growth is nothing but some analytic continuation from real to imaginary 
time of a quantum quench in the Lieb-Liniger model. 

Despite this very clear correspondence between the two problems, there has not been 
much cross fertilisation of ideas, techniques, and results between the two fields that have been evolving  
mostly independently. 
In this short paper we begin filling this gap. We show how recent quantum-quench results by 
De Nardis et al. \cite{nwbc-13} can be adapted to partially re-obtain the KPZ growth starting 
from a flat initial condition, a problem solved  
by means of the Bethe ansatz in \cite{we-flat,we-flatlong}.
In Ref. \cite{nwbc-13}, the authors studied the non-equilibrium dynamics of a gas of $n$ interacting bosons 
on a one-dimensional ring of perimeter $L$ after a quantum quench. The initial state was fixed to be the non-interacting 
Bose Einstein condensate (BEC), 
and arbitrary repulsive interactions were turned on for times $t>0$ in the form of a Lieb Liniger model. 
To determine the time evolution for $t>0$, they conjectured and motivated a formula for the overlap 
between arbitrary Bethe eigenstates with real rapidities and the uniform BEC state 
(we have been made aware that a proof of this formula has been recently achieved \cite{proof}). 
This formula was used to characterise the stationary values of local observables \cite{nwbc-13}. 
As stressed in Ref. \cite{we-flatlong} (but for the Bethe eigenstates of the attractive model and hence with
complex rapidities), 
these overlaps are also the starting point for the solution of the KPZ equation with flat initial condition.
The lack of a closed formula for the overlaps at fixed system size $L$, was overcome in \cite{we-flat} by 
working directly in the infinite space $L=+\infty$, but introducing appropriate regularisations of the initial condition.
The knowledge of the overlap can shed some new light for this 
problem and should help rewriting the solution of the KPZ problem in a more direct way. 
Unfortunately at present the overlaps are know only in the repulsive regime and for an even number of 
particles, while for the KPZ problem we need to consider attractive interactions and arbitrary integer number of particles $n$. 
In this manuscript we show how to generalise the formula for the overlap to a subset of the 
eigenstates of the attractive problem 
recovering some of the results of \cite{we-flat,we-flatlong} by a completely different method.
We have not yet been able to generalise the fixed size overlap formula to odd number $n$ of particles.
Also for even $n$ there are eigenstates of the attractive problem 
for which the fixed size overlap formula is hard to apply. 
However, we can proceed by reverse engineering and use the known results for the 
KPZ flat initial condition problem \cite{we-flat, we-flatlong} to conjecture a formula for the 
overlap between the BEC and an arbitrary string state,
which is valid at arbitrary fixed $n$ and 
large $L$, i.e. in the limit of zero density of particles (which is the KPZ working scheme).

The manuscript is organised as following.
In Sec. \ref{Sec:LL} we recall the Bethe ansatz solution of the Lieb Liniger model both for repulsive and 
attractive interaction. 
In Sec. \ref{Sec:ov1} we recall the overlap formula of Ref. \cite{nwbc-13}
and generalise it to complex rapidities.
In Sec. \ref{Sec:ov2} we take explicitly the thermodynamic limit of this overlap on string solutions 
and conjecture the other overlaps we need for the KPZ problem. 
In Sec. \ref{Sec:KPZ} we apply all the previous results to the KPZ equation
and finally in Sec. \ref{concl} we draw our conclusions and discuss many open issues. 
Two appendices report some technical details.

\section{Lieb Liniger model}
\label{Sec:LL}

We consider here the Lieb-Liniger Hamiltonian in Eq. (\ref{LL}) for $n$ bosons with interaction parameter $c$
which is solvable by means of the Bethe ansatz \cite{ll}. 
The symmetric eigenstates $|\mu \rangle$ of $H_n$ are indexed by a
set of all distinct rapidities $\{\lambda_\alpha\}_{\alpha=1,..n}$ and their many-body wave-functions 
$\Psi_\mu(x_1,..x_n)= \langle x_1,..x_n |\mu \rangle$ take the form \cite{ll} 
\be \label{def1}\fl
\Psi_\mu(x_1,..x_n) =  \sum_P A_P \prod_{j=1}^n e^{i \sum_{\alpha=1}^n \l_{P_\alpha} x_\alpha} \, , \quad 
A_P=\prod_{n \geq \beta > \alpha \geq 1} \Big(1- \frac{i c ~\sgn(x_\beta - x_\alpha)}{\lambda_{P_\beta} - \lambda_{P_\alpha}}\Big)\,.
\ee
The sum runs over all $n!$ permutations $P$ of the rapidities $\l_\alpha$.
As written, these Bethe wave functions are not normalized and satisfy 
\be 
\langle x \cdots x |\mu\rangle =  \Psi_\mu(x,..x) = n! e^{ i x \sum_\alpha \lambda_\alpha} \,.
\label{Psixeq}
\ee 

We work here with the model defined on a circle of period $L$, i.e. $ \Psi_\mu(x_1,..x_n)$ must be periodic in any of the variables 
with period $L$, which implies that the set of rapidities $\{\lambda_\alpha\}_{\alpha=1,..n}$ must be solution to the Bethe equations 
\begin{eqnarray}
e^{i \lambda_{\alpha} L} = \prod_{\beta \neq \alpha} \frac{\lambda_{\alpha} - \lambda_{\beta} + i c}
{\lambda_{\alpha} - \lambda_{\beta} - i c},
\label{BE}
\end{eqnarray}
The corresponding eigenstates are called the Bethe eigenstates. 

The state in Eq. (\ref{def1}) is not normalised and its norm is given by the celebrated 
Gaudin-Korepin formula \cite{gk}
for the norm of a Bethe eigenstate which, in the present conventions,  reads
\be
 || \mu||^2 = n!  \prod_{1 \leq \alpha < \beta \leq n} \frac{(\lambda_\alpha-\lambda_\beta)^2 + c^2}{(\lambda_\alpha-\lambda_\beta)^2} \det  G, 
 \label{Gaudin}
\ee
where $G$ is the $n$-by-$n$ Gaudin matrix with elements
\bea
&& G_{\alpha \beta} = \delta_{\alpha \beta} (L + \sum_{\gamma=1}^n K(\lambda_\alpha - \lambda_\gamma)) - K(\lambda_\alpha - \lambda_\beta), \\
&& K(x) = \frac{2 c}{x^2 + c^2}. 
\eea 

For repulsive interactions $c>0$, it turns out that all the solutions to the Bethe equations (\ref{BE}) are real \cite{ll,ll2}. 
The situation dramatically changes when switching the sign of $c$ to attractive 
interactions $c= - \bar c<0$ because there are solutions to the Bethe equations which are 
complex \cite{m-65} as recalled in the following subsection. 

\subsection{Attractive Lieb-Liniger and string solutions}

For $c<0$ and in the limit $L\to\infty$, the rapidities solutions to the Bethe equations arrange in {\it strings}.
A general eigenstate is built by partitioning the $n$ particles into a set of $1 \leq n_s \leq n$ 
strings formed by $m_j \geq 1$ particles with $n=\sum_{j=1}^{n_s} m_j$. 
The rapidities associated to these states are written as
\be\label{stringsol}
\l^{j, a}=k_j +\frac{i\cb}2(m_j+1-2a)+i\d^{j,a}.
\ee 
Here, the index $a = 1,...,m_j$ labels the rapidities within the $j$-th string with $j=1,\dots n_s$.  
The $\d^{j,a}$ are deviations which fall off exponentially with system size $L$.
As it should be clear, perfect strings (i.e. with $\d=0$) are exact eigenstates in the limit $L\to\infty$ for arbitrary $n$.
The string states have momentum  and energy 
\be
K_\mu=\sum_{j=1}^{n_s} m_j k_j,\qquad E_\mu=\sum_{j=1}^{n_s} \left(m_j k_j^2-\frac{\cb^2}{12} m_j(m_j^2-1)\right).
\label{enmom}
\ee 

The general formula for the norm of a string state is obtained by inserting the string solutions (\ref{stringsol}) 
into Eq. (\ref{Gaudin}) and then carefully taking the limit
$L \to \infty$, as shown in Ref. \cite{cc-07} with the final result
\be
||\mu||^2 =  \frac{n!  (L \bar{c})^{n_s} } {\bar{c}^{n}} \prod_{j=1}^{n_s} m_j^{2} 
\prod_{1\leq i<j\leq n_s}  \frac{4(k_i-k_j)^2 +(m_i+m_j)^2 c^2}{4(k_i-k_j)^2 +(m_i-m_j)^2 c^2}. \label{norm} 
\ee

\section{Quantum quenches and the overlap formula} 
\label{Sec:ov1}

A quantum quench \cite{cc-06,revq} is a non-equilibrium protocol in which a Hamiltonian parameter is
suddenly changed. 
A very important example of quantum quenches is represented by a change of the interaction 
parameter in the Lieb-Liniger Hamiltonian (\ref{LL}), i.e. the evolution from 
the ground-state of the Hamiltonian (\ref{LL}) with an arbitrary interaction $c_0$ and governed by 
the same Hamiltonian with a different $c\neq c_0$. 
Despite of the integrability of the Hamiltonian before and after the quench, 
and despite the many investigations \cite{cro,grd-10,mc-12,jap,a-12,ce-13,ksc-13,kcc-13,m-13,nwbc-13}, 
the solution of this (only apparently simple) problem is still well beyond the reach of 
the most advanced mathematical tools of Bethe ansatz. 

The most general way to tackle a quantum quench problem is to write the initial state $|\Phi_0\rangle$
in terms of the eigenstates of the Hamiltonian $H$ governing the time evolution. Let us for the moment generically denote these eigenstates 
as $|n\rangle$ (they will be denoted as $|\mu\rangle$ below in the case of the Lieb Liniger model). We have:
\be
|\Phi_0\rangle=\sum_n a_n |n\rangle,
\ee
where $a_n$ represent the overlaps $a_n\equiv \langle n|\Phi_0\rangle$ between the initial state and the 
eigenstates.
Once the overlaps are known, the time evolved state is simply
\be
|\Phi(t)\rangle=\sum_n a_n e^{-i E_n t}|n\rangle,
\ee
$E_n$ being the energy of the state $|n\rangle$.
This provides the time dependent expectation values of an observable $O$ as
\be
\langle \Phi(t)|O |\Phi(t)\rangle= \sum_{nm} a_n a^*_m e^{i(E_m- E_n) t} \langle m| O |n\rangle,
\ee
once the form factors $\langle n|O|m\rangle$ are known. Calculating the form factors 
(of the most relevant local operators) is the main goal of the algebraic Bethe ansatz and 
quantum inverse scattering program \cite{KorepinBOOK}, which for the Lieb-Liniger model has been practically 
completed \cite{s-89,kks-97,cc-06b,kmt-09}. 
However, it is still impossible to calculate the overlap $a_n$ for an arbitrary quench in the 
Lieb-Liniger model and for this reason in Refs. \cite{grd-10,ksc-13,kcc-13,nwbc-13} the simplest initial state has been considered, 
namely the ground state of the Hamiltonian with $c_0=0$, i.e. the BEC state with a uniform wave function 
$\Phi_0(x_1,..x_n)={\rm const}$. 

In a recent work De Nardis et al. \cite{nwbc-13} conjectured a formula for the overlap between the BEC state and a Bethe state
in the case of an 
{\it even number of particles $n$}, a conjecture that they argue is valid for $c>0$ and real rapidities. 

First they state that only ``parity invariant'' eigenstates have a non-zero overlap with the BEC state.
It is easy to see on (\ref{def1}) that the parity transformation $x_j \to - x_j$ transforms the
Bethe state with a set of rapidities $\{ \lambda_\alpha \}$ into the set $\{ - \lambda_\alpha \}$. Hence
parity invariant states are such that:
\bea \label{parityinv} 
&& \lambda_{\alpha+n/2} = - \lambda_\alpha \quad , \quad \alpha=1,\dots, \frac{n}{2}  \quad , \quad n~~ {\rm even} \\
&& \lambda_{\alpha+(n-1)/2} = - \lambda_\alpha \quad , \quad \alpha=1,\dots, \frac{n-1}{2} \quad , \quad  \lambda_n=0 
\quad , \quad n ~~ {\rm odd} \nn
\eea
i.e. each rapidity has a partner with opposite sign, unless vanishing. Recall that all rapidities
must be different, hence only one zero rapidity is allowed (and only for $n$ odd). 
In \cite{we-flat, we-flatlong} (Section 4) in the KPZ context, we had observed that if no partial sum of rapidities vanish 
then the overlap is zero. In addition, any state found to contribute was indeed of the parity invariant form, leading
to the pairing mechanism discussed below. A more systematic rationale is given in \cite{nwbc-13}. It invokes the existence 
of an infinite set of conserved charges $Q_p$ , $p=1,2..$, for the Lieb Liniger 
model, i.e. mutually commuting operators, and commuting with the Hamiltonian $H_n$, which stems from its integrability property.
Their general expression is simple in the Bethe state basis,
$Q_p|\mu\rangle=\sum_\alpha \lambda_\alpha^p|\mu\rangle$, but  
in terms of position/momentum only the lowest charges $Q_0$ (particle number), $Q_1$ (total momentum), and $Q_2$ (total energy) 
are simple, further explicit expressions being known \cite{conserved} only for $p=3,4$. 
In Ref. \cite{nwbc-13} it was argued that 
\footnote{this can be checked for the first quantised form of $Q_3(\{x_\alpha,p_\alpha\})$ given in \cite{conserved}, and holds if in all (unknown) $Q_{2 m+1}(\{x_\alpha,p_\alpha\})$ 
the derivatives $p_j=-i \partial_{x_j}$ can be put to the right. In the second quantised form it is also natural since 
$(\Psi \Psi^+)^m$ is the only term not containing derivatives which has a non-zero expectation on a uniform state
see e.g. \cite{ksc-13}. 
} 
\be
Q_{2m+1} |\Phi_0\rangle = 0 ,
\ee 
acting on the (uniform) BEC state. This implies that the expectation between the BEC state and an 
arbitrary Bethe state, is $0=\langle \Phi_0 |Q_{2m+1} |\mu \rangle= \langle \Phi_0 |\mu \rangle \sum_j \lambda_j^{2m+1}$.
This guarantees that the overlap is zero, unless $\sum_j \lambda_j^{2m+1}=0$ for all $m$, i.e. each rapidity
must have a partner with opposite sign. It is evident that this argument is still valid even if we allow
the rapidities to be complex numbers as in the case of the attractive Bose gas. Note that this property
is obtained for the uniform initial state  $\Phi_0(x_1,..x_n)={\rm const}$, and that parity and global (i.e. center of mass) 
translational invariance of the 
initial state are not sufficient to guarantee this property.

For the non-zero overlaps, 
the conjecture of Ref. \cite{nwbc-13}, in our convention (i.e. unnormalised initial wave function $\Phi_0(x_1,..x_n)=1$
and unnormalised Bethe state $|\mu\rangle$ as in Eq. (\ref{def1})), 
reads
\bea \label{overlap} \fl
&& \langle \Phi_0| \mu\rangle= \prod_{\alpha=1}^n \int_0^L dx_\alpha 
 \Psi_\mu(x_1,..x_n) \\
 && = n!  \prod_{\alpha=1}^{n/2} \frac{1}{\lambda_\alpha \sqrt{\frac{\lambda_\alpha^2}{c^2} + \frac{1}{4}}} 
\prod_{1 \leq \alpha < \beta \leq n} \frac{\sqrt{ (\lambda_\alpha-\lambda_\beta)^2 + c^2}}{|\lambda_\alpha-\lambda_\beta|} 
\times \det G^Q, 
\eea
where $G^Q$ is a modified Gaudin $n/2\times n/2$ matrix with elements
\bea
&& G^Q_{\alpha \beta} = \delta_{\alpha \beta} 
(L + \sum_{\gamma=1}^{n/2} K^Q(\lambda_\alpha,\lambda_\gamma)) - K^Q(\lambda_\alpha,\lambda_\beta), \\
&& K^Q(x,y) = K(x-y) + K(x+y), 
\eea 
taking explicitly into account the parity invariant condition. 
In the above formula the $\lambda_\alpha$, $\alpha=1,..n/2$ are the $n/2$ positive rapidities 
and we recall that the number of particles is even.

\subsection{Generalisation of the overlap formula to complex rapidities}

We first note that Eq. (\ref{overlap}) can be extended to arbitrary complex rapidities satisfying the Bethe equations. 
To this goal we rewrite it as an analytic function of all $\lambda$s, that, taking into account the parity invariance, is
\bea \label{overlap2}
&& \fl  \langle \Phi_0| \mu\rangle = n! c^{n/2} \prod_{\alpha=1}^{n/2} \frac{1}{\lambda_\alpha^2}  
\prod_{1 \leq \alpha < \beta \leq n/2}  \frac{(\lambda_\alpha-\lambda_\beta)^2 + c^2}{(\lambda_\alpha-\lambda_\beta)^2} 
\frac{(\lambda_\alpha+\lambda_\beta)^2 + c^2}{(\lambda_\alpha+\lambda_\beta)^2} \times \det G^Q .
\eea 
We have checked that this equation holds for arbitrary complex rapidities for $n=2,4,6$.
The explicit closed expressions for the overlaps for low number of particles, obtained by brute force integration 
of the Bethe wave functions, are given in  \ref{AppA} and they agree with the above general formula.
We then conjecture that Eq. (\ref{overlap2}) is valid for a generic Bethe state of even number of particles 
with paired rapidities and arbitrary sign for $c$.

\section{Overlap formula for string states}
\label{Sec:ov2}

We will now study the overlap formula in Eq. (\ref{overlap2})
in the limit $L\to\infty$ at fixed $n$ when the solutions to Bethe
equations organise into string states. 
First, let us notice that the most general parity invariant string state can be constructed by splitting the 
$n_s$ strings into $N$ pairs of strings with opposite momenta and $M$ single strings of zero momentum 
and clearly $n_s=2 N+M$ ($N$ and $M$ are the same of Refs. \cite{we-flat,we-flatlong}). 
Although states with $M>1$ seem naively ruled out by the rule that all rapidities should be distinct,
we must remember that there is a limiting procedure as $L \to +\infty$. It turns out,
as discovered in Refs. \cite{we-flat,we-flatlong} and discussed below, that the
``anomalous" states with $M>1$ are possible in that limit and important (see also Ref. \cite{hc-07} 
for a discussion of deformed strings in spin chains). 
Let us now consider the various cases separately.

\subsection{Paired string states, $M=0$}

We start with an eigenstate $|\mu\rangle$ of $n_s$ strings, such that all strings are paired.
In that case $n_s=2 N$ is even, and $M=0$. We can label their momenta $k_1,-k_1,k_2,-k_2,..k_p,-k_p,..$ 
with $p=1,\dots N$ and denote their
sizes as $n_1,n_1,n_2,n_2,..$. With no loss of generality we can take all $k_i>0$. 
From (\ref{norm}) the norm of this state is
\bea
\label{norm2N}
\fl  ||\mu||^2 &=& \frac{n!  (L \bar{c})^{n_s} } {\bar{c}^{n}} \prod_{p=1}^{N} n_p^{4} 
\frac{4 k_p^2 +n_p^2 c^2}{4k_p^2 } \nonumber \\
 \fl && \times 
\prod_{1\leq p<q\leq N} \Big[  \frac{4(k_p-k_q)^2 +(n_p+n_q)^2 c^2}{4(k_p-k_q)^2 +(n_p-n_q)^2 c^2} 
\frac{4(k_p+k_q)^2 +(n_p+n_q)^2 c^2}{4(k_p+k_q)^2 +(n_p-n_q)^2 c^2}  \Big]^2 ,
\eea

Let us now examine the overlap formula (\ref{overlap2}). 
In the limit $L \to \infty$ one can repeat the analysis of Ref. \cite{cc-07} (Appendix B there) used to evaluate $\det G$ 
on a string state in order to calculate $\det G^Q$.
It is straightforward to convince ourselves that step by step the new calculation 
parallels the one for $\det G$ in \cite{cc-07}. 
Indeed, the kernel $K^Q(\lambda_\alpha,\lambda_\beta)$ has exactly the same divergences in the string deviations
as $K(\lambda_\alpha-\lambda_\beta)$. 
Hence $\det G^Q$ is related to the norm of the ``half" Bethe state, which we call $|\mu'\rangle$, i.e. the state with only 
$N$ strings (say the ones with positive momenta $k_p$, $p=1,\dots, N$) and $n/2$ particles. 
Let us order the rapidities such that $\lambda_\alpha$ with $\alpha \leq n/2$ belong to $|\mu'\rangle$. 
Thus, in the limit $L\to\infty$ we obtain
\bea
\lim_{L \to \infty}  
\prod_{1 \leq \alpha <\beta \leq n/2} \frac{(\lambda_\alpha-\lambda_\beta)^2 + c^2}{(\lambda_\alpha-\lambda_\beta)^2} \det G^Q 
= \frac{1}{(n/2)!} ||\mu'||^2.
\eea 
Hence we find in the limit $L\to\infty$
\bea \label{overlap3}
&& \fl  \langle \Phi_0| \mu\rangle = n! c^{n/2} \prod_{\alpha=1}^{n/2} \frac{1}{\lambda_\alpha^2}  
\prod_{1 \leq \alpha < \beta \leq n/2} 
\frac{(\lambda_\alpha+\lambda_\beta)^2 + c^2}{(\lambda_\alpha+\lambda_\beta)^2} \frac{1}{(n/2)!} ||\mu'||^2.
\eea 

Let us now rewrite the various pieces in the overlap above. 
The norm of $|\mu'\rangle$ is
\be
 ||\mu'||^2 =  \frac{(n/2)!  (L \bar{c})^{N} } {\bar{c}^{n/2}} \prod_{p=1}^{N} n_p^{2} 
\prod_{1\leq p<q\leq N}  \frac{4(k_p-k_q)^2 +(n_p+n_q)^2 c^2}{4(k_p-k_q)^2 +(n_p-n_q)^2 c^2}.
\ee
The product over the rapidities can be split in an intra-string contribution
\be
\prod_{a=1}^{n_p} \frac{1}{\lambda_{p,a}^2}  
\prod_{1 \leq a < b \leq n_p} 
\frac{(\lambda_{p,a}+\lambda_{p,b})^2 + c^2}{(\lambda_{p,a}+\lambda_{p,b})^2} = 2^{2 n_p} 
\prod_{q=0}^{n_p-1} \frac{1}{4 k_p^2 + q^2 c^2}, 
\ee
and an inter-string contribution
\be
\prod_{a=1}^{n_p} \prod_{b=1}^{n_q} 
\frac{(\lambda_{p,a}+\lambda_{q,b})^2 + c^2}{(\lambda_{p,a}+\lambda_{q,b})^2} = 
\frac{4 (k_p+k_q)^2 + c^2 (n_p+n_q)^2}{4 (k_p+k_q)^2 + c^2 (n_p-n_q)^2}.
\ee
both being calculated inserting (\ref{stringsol}). 

Putting together the three contributions, we obtain our final formula for the overlap 
for parity invariant string states consisting of $2N$ strings\footnote{ This formula reproduces the results of Ref. \cite{we-flatlong} 
by direct calculation in the cases $n=2$, $N=1$, (50) there, and $n=4$, $N=1,2$, 
(B.2) and (B.4) there.}
\bea
 \fl  \langle \Phi_0| \mu\rangle &=& n! (-1)^{n/2}   (L \bar{c})^{N}   \prod_{p=1}^N 2^{2 n_p} n_p^{2}  
\prod_{q=0}^{n_p-1} \frac{1}{4 k_p^2 + q^2 c^2} \nonumber\\
\fl &&  
\times \prod_{1 \leq p < q \leq N} 
\frac{4 (k_p+k_q)^2 + c^2 (n_p+n_q)^2}{4 (k_p+k_q)^2 + c^2 (n_p-n_q)^2}  \frac{4(k_p-k_q)^2 +(n_p+n_q)^2 c^2}{4(k_p-k_q)^2 +(n_p-n_q)^2 c^2}.
\label{ov2N}
\eea 
Notice that, in our conventions, the overlap has dimension $\bar c^{-n}$, where $\bar c$ has dimension 
inverse of length (i.e. same dimension as a momentum or a rapidity).

We need to mention now that in order to get this result we strongly used the parity invariance 
of the state also at finite $L$, i.e. also the string deviations are parity invariant, in the sense 
that the symmetry $\lambda\to-\lambda$ is exact in the complex plane.

\subsection{One zero momentum string with even number of particles}

Another case that can be handled starting from Eq. (\ref{overlap2}) is that of a single zero momentum string 
composed of $m$ particles with $m$ even, which is clearly parity invariant.  
We fix the first $m/2$ rapidities (entering in $G^Q$) to be the ones with positive imaginary part.
Also in this case, the calculation of $\det G^Q$ parallels  
the one for $\det G$ in \cite{cc-07} because 
the kernel $K^Q(\lambda_\alpha,\lambda_\beta)$ has  the same divergences in the string deviations
as $K(\lambda_\alpha-\lambda_\beta)$. 
Thus, following the calculation in Ref. \cite{cc-07}, in the limit $L\to\infty$ we obtain
\bea
\lim_{L \to \infty}  
\prod_{1 \leq \alpha <\beta \leq m/2} \frac{(\lambda_\alpha-\lambda_\beta)^2 + c^2}{(\lambda_\alpha-\lambda_\beta)^2} \det G^Q 
= \frac{L \cb}{\cb^{m/2}} \left(\frac{m}{2}\right)^2.
\eea 
Hence we have in the limit $L\to\infty$
\be \label{overlap4}
   \langle \Phi_0| \mu\rangle = m! c^{m/2} \prod_{\alpha=1}^{m/2} \frac{1}{\lambda_\alpha^2}  
\prod_{1 \leq \alpha < \beta \leq m/2} 
\frac{(\lambda_\alpha+\lambda_\beta)^2 + c^2}{(\lambda_\alpha+\lambda_\beta)^2} \frac{L \cb}{\cb^{m/2}} \frac{m^2}4.
\ee
The calculation of the remaining product is even easier than in the previous case, because we have only an
intra-string contribution and so 
\be
\prod_{a=1}^{m/2} \frac{1}{\lambda_{a}^2}  
\prod_{1 \leq a < b \leq m/2} 
\frac{(\lambda_{a}+\lambda_{b})^2 + c^2}{(\lambda_{a}+\lambda_{b})^2} = 
\frac{(-1)^{m/2}}{c^m} \frac{2^{m+1}}{m!}, 
\ee
leading to 
\be
\langle \Phi_0| \mu\rangle = \frac{L}{\cb^{m-1}} 2^{m-1}m^2. 
\ee
This agrees with Eqs. (50) and (B.1) in Ref. \cite{we-flatlong} for $m=2,4$ and with Eq. (60) for general $m$ (including odd).

Clearly, we can easily now handle the case with $2N$ paired strings (each with $n_p$ particles with $p=1\dots N$) 
and a single string with zero momentum 
and even number of particles combining the calculation above with the one in the previous subsection. 
We just report the final result 
\bea \fl    \langle \Phi_0| \mu\rangle
&=& \frac{n!  (L \bar{c})^{N+1} } {\bar{c}^{n}} \frac{2^{m-1}m^2}{m!} 
 \prod_{p=1}^{N}  (-1)^{n_p}  2^{2 n_p}n_p^2 \frac{4 k_p^2 + c^2(n_p+m)^2}{4 k_p^2 + c^2(n_p-m)^2}
 \prod_{q=0}^{n_p-1} \frac{1}{4 k_p^2/c^2 + q^2}  
  \nonumber \\
 \fl && \times
\prod_{1\leq p<q\leq N} \Big[  \frac{4(k_p-k_q)^2 +(n_p+n_q)^2 c^2}{4(k_p-k_q)^2 +(n_p-n_q)^2 c^2} 
\frac{4(k_p+k_q)^2 +(n_p+n_q)^2 c^2}{4(k_p+k_q)^2 +(n_p-n_q)^2 c^2}  \Big].  
\end{eqnarray}

\subsection{More general string states}

Let us now consider the general state with $n_s = 2 N + M$ strings. Its norm is given from Eq. (\ref{norm}) by
\bea
\fl  ||\mu||^2 &=& \frac{n!  (L \bar{c})^{n_s} } {\bar{c}^{n}} \prod_{j=1}^{M} m_j^{2} 
\prod_{1\leq i<j\leq n_s}  \frac{(m_i+m_j)^2}{(m_i-m_j)^2} 
\prod_{p=1}^{N} n_p^{4} 
\frac{4 k_p^2 +n_p^2 c^2}{4k_p^2 }  \nonumber \\
 \fl && \times \prod_{1 \le p \le N,1 \le j \le M} \Big[\frac{4 k_p^2 + c^2(n_p+m_j)^2}{4 k_p^2 +c^2 (n_p-m_j)^2}\Big]^2 \nonumber \\
 \fl && \times
\prod_{1\leq p<q\leq N} \Big[  \frac{4(k_p-k_q)^2 +(n_p+n_q)^2 c^2}{4(k_p-k_q)^2 +(n_p-n_q)^2 c^2} 
\frac{4(k_p+k_q)^2 +(n_p+n_q)^2 c^2}{4(k_p+k_q)^2 +(n_p-n_q)^2 c^2}  \Big]^2 ,
\label{norm3}
\eea 
which clearly reduces to Eq. (\ref{norm2N}) for $M=0$.

There are however a series of technical difficulties in order to apply the overlap formula to strings
of zero momentum, in particular because two rapidities can have the same value.
For finite $L$, this problem is overcome by string deviations which are different for particles 
belonging to different strings. 
Thus in order to have the correct result in an infinite system, one should 
first consider finite $L$ taking into account the string deviations, then calculate the limit of the overlap using 
the Bethe equations and finally take the limit $L\to\infty$. 
An explicit example of this procedure for the ``anomalous'' string $M=2$ with $m_1=1$ and $m_2=3$
is reported in \ref{AppB}.
We call this string anomalous because there are two equal rapidities (both equal to zero) for $L=\infty$.
States like this one are known to have no weight in equilibrium problems (see e.g. \cite{cc-07}), 
and so they are usually ignored. 
The result for the overlap in \ref{AppB} shows instead that out of equilibrium they are very important
and they should be properly taken into account, as  done for the KPZ equation \cite{we-flat,we-flatlong}
(see also below).

We have not been able to derive a formula for the overlap of a generic string state starting 
from the result of De Nardis et al. (\ref{def1}), also, but not only,  because there is no equivalent conjecture 
for an odd number of particles. 
However, we can arrive at a conjecture for the infinite volume overlap from the following
heuristic argument, after realising the 
recursive hierarchical structure of the formula for the norm (\ref{norm3})
when paired strings go to zero momentum.
Indeed, let us start from the norm of paired string in Eq. (\ref{norm2N}).
When one momentum $k_p$ goes to zero, we would have two strings of $n_p$ particles with zero momentum. 
This is clearly not allowed and indeed the corresponding norm diverges like $k_p^{-2}$.
However the quantity $k_p^2 ||\mu||^2$ has a finite limit and, by simple inspection, it is equal to the 
norm of a state with $2(N-1)$ paired strings times the square of the factor in the second line of Eq. (\ref{norm3})  
(times some other trackable factors). 
This reflects the quasi-free nature of strings in infinite volume, because it is like having an (impossible) state with
two equal strings with zero momentum.
Thus, we expect that taking the square root of the terms involving the two strings with zero momentum, 
we would get the norm of the state with  $2(N-1)$ paired strings and one string of zero momentum, 
as indeed it happens. 
Repeating this procedure for $M$ strings, one arrive from Eq. (\ref{norm2N}) to the general norm
(\ref{norm3}).
Notice that this recursive structure is valid {\it only} in the infinite volume limit. 

Having understood this, it is reasonable to expect that the same recursive hierarchical structure
of the norm must be valid also for the overlap, but with the additional problem of fixing a sign when 
taking the square root (by definition the norm square is positive, but not the overlap). 
Thus starting from Eq. (\ref{ov2N}), letting $M$ momentum to zero, we have a conjecture for the 
overlap apart from a sign. The final structure we obtain is equivalent to what has been obtained for KPZ 
equation \cite{we-flat,we-flatlong} (with a regularised initial state).
Thus, taking the correct sign from Refs. \cite{we-flat,we-flatlong} we end up with the following conjecture 
for the overlap 
\bea \fl   \langle \Phi_0| \mu\rangle
&=& \frac{n!  (L \bar{c})^{M+N} } {\bar{c}^{n}}
\frac{2^{n} }{2^{M}}    \prod_{j=1}^M \frac{m_j^2}{m_j!} \prod_{1 \leq i<j \leq M} \!\! (-1)^{\min(m_i,m_j)} \frac{m_i+m_j}{|m_i - m_j|} 
  \nn \\
 \fl && \times  \prod_{p=1}^{N}  (-1)^{n_p}  n_p^2 \prod_{q=0}^{n_p-1} \frac{1}{4 k_p^2/c^2 + q^2 }  \prod_{1 \le p \le N,1 \le j \le M} 
 \Big[\frac{4 k_p^2 + (n_p+m_j)^2 c^2}{4 k_p^2 + (n_p-m_j)^2 c^2}\Big] \nonumber \\
 \fl && \times
\prod_{1\leq p<q\leq N} \Big[  \frac{4(k_p-k_q)^2 +(n_p+n_q)^2 c^2}{4(k_p-k_q)^2 +(n_p-n_q)^2 c^2} 
\frac{4(k_p+k_q)^2 +(n_p+n_q)^2 c^2}{4(k_p+k_q)^2 +(n_p-n_q)^2 c^2}  \Big].  
\end{eqnarray}
Note that this formula is more than simply a conjecture, since it is also a corollary of the exact but quite indirect derivation 
presented in Refs. \cite{we-flat,we-flatlong}, as discussed again below. Our present aim was to arrive at this formula 
directly from the finite $L$ overlap conjecture of de Nardis et al., via some heuristic arguments.

We stress once again that this overlap is expected to be valid for an arbitrary string state, but only in the 
infinite volume limit for finite $n$, i.e. in the limit of zero density. 
While this limit is of little physical interest for the repulsive Bose gas, in the attractive regime is the 
correct one for KPZ equation \cite{we-flat} and also for the Bose gas \cite{cc-07}.

It is worth mentioning that the (dimensionless) normalised overlap turns out to have a very  simple form
\bea
 \fl    \frac{\langle \Phi_0| \mu\rangle}{||\mu|| ||\Phi_0||} 
&=& \sqrt{n!} (L \bar{c})^{\frac{M-n}{2}} 
\frac{2^{n} }{2^{M}}    \prod_{j=1}^M \frac{1}{(m_j-1)!} \prod_{1 \leq i<j \leq M} \!\! (-1)^{\min(m_i,m_j)}
  \nn \\
 \fl && \times  \prod_{p=1}^{N}  (-1)^{n_p}  \prod_{q=0}^{n_p-1} \frac{1}{4 k_p^2/c^2 + q^2}  \sqrt{\frac{4 k_p^2}{4 k_p^2 + n_p^2 c^2}}.
\end{eqnarray}
using $||\Phi_0||=L^{n/2}$. 
This is the quantity of major interest for quantum quench problems. 
Instead, for the product needed for the KPZ equation we have
\bea
&& \fl  \frac{\langle \Phi_0| \mu\rangle}{||\mu||^2}   
= \frac{2^{n} (\bar{c} L)^{-N}  }{2^{M}}   \prod_{j=1}^M \frac{1}{m_j!} 
\prod_{1 \leq i<j \leq M} \!\! (-1)^{\min(m_i,m_j)} \frac{|m_i - m_j|}{m_i+m_j} \nn
\\
\fl && \times \prod_{p=1}^{N}    \frac{(-1)^{n_p} }{n_p^2} \prod_{p=1}^N \prod_{q=1}^{n_p} \frac{1}{4 k_p^2/c^2 + q^2} 
\prod_{1 \le p \le N,1 \le j \le M} \frac{4 k_p^2 + (n_p-m_j)^2}{4 k_p^2 + (n_p+m_j)^2}
  \nn 
  \\
   \fl  && \times  
   \prod_{1\leq p<q\leq N}
\frac{4 (k_p-k_q)^2 +(n_p-n_q)^2}{4 (k_p-k_q)^2 +(n_p+n_q)^2}
\frac{4 (k_p+k_q)^2 +(n_p-n_q)^2}{4 (k_p+k_q)^2 +(n_p+n_q)^2}  ,
\label{ovnorm2}
\end{eqnarray}
which provides the same results for KPZ generating function in Refs. \cite{we-flat,we-flatlong}, as we shall discuss in next section.

\section{Application to KPZ equation}
\label{Sec:KPZ}

The KPZ equation is an equation describing the non-equilibrium
growth in time $t$ of an interface of height $h(x,t)$ in the presence
of noise \cite{KPZ}. 
It defines a universality class encompassing numerous models and physical systems \cite{kpzreviews}.
In one dimension, i.e. $x \in R$, it reads
\be \label{kpzeq}
\partial_t h = \nu \nabla^2 h + \frac{1}{2} \lambda_0 (\nabla h)^2 + \sqrt{D} \eta(x,t)\,,
\ee
where $\overline{\eta(x,t) \eta(x,t')}=\delta(x-x') \delta(t-t')$ is a centered Gaussian white noise. 
By rescaling of space and time it can be brought to the choice \cite{we,we-flatlong,ld-14}  of parameters
$\nu=1$, $\lambda_0=2$ and $D=2 \bar c$ 
which we use from now on. 

The solution of the KPZ equation with flat initial condition $h(x,t=0)=0$, using the Cole-Hopf 
transformation, can be written as 
\bea
h(x,t) = \ln Z(x,t) ,
\eea 
where $Z(x,t)$ is the partition sum of the directed polymer with one end at $x$ and one free end
\bea
Z(x,t) = \int  dy Z(xt|y0). 
\eea 
where, for the KPZ problem on the infinite line, the integration runs for $y \in ]-\infty,+\infty[$. 

To obtain the one point probability distribution function of $Z(x,t)$ one can set $x=0$. Furthermore here we consider the
problem on a circle of size $L$ with periodic boundary condition. Hence we define
\bea
Z_{\rm flat}(t) =  \int _0^L dy Z(0t|y0) ,
\eea  
and study the integer moments $\overline{Z_{\rm flat}^n}$ which, by definition, in the replica approach are
\be
\overline{Z_{\rm flat}^n(t)} =  \big( \prod_{j=1}^n  \int_{0}^L dy_j \big) \langle y_1 \dots y_n |e^{- t H_n} |0 \dots 0 \rangle \,,
\ee
where $H_n$ is the Lieb Liniger Hamiltonian (\ref{LL}).

Assuming completeness of the string states, we can insert a resolution of the identity over the Bethe states 
$|\mu\rangle$ (representing all the possible allowed rapidities) and obtain 
\bea
\overline{Z_{\rm flat}^n(t)}&=& \sum_\mu   \big( \prod_{j=1}^n \int_{0}^L dy_j \big)
\frac{ \langle y_1 \dots y_n |\mu\rangle \langle\mu| 0 \dots 0 \rangle}{||\mu ||^2} e^{-t E_\mu}\nn \\&=&
\sum_\mu \frac{\Psi_\mu^*(0,..0)}{||\mu ||^2}   \langle \Phi_0| \mu\rangle e^{-t E_\mu}
\,,
 \label{sumf}
\eea
where $E_\mu$ is the energy of the state $\mu$ (cf. Eq. (\ref{enmom})) and in the last line we recognized the definition of the overlap 
$\langle \Phi_0| \mu\rangle=  \Big(\prod_{j} \int_{0}^L dy_j\Big) \Psi_\mu(y_1\dots y_n)$.
Note that in our normalisation $\Psi_\mu^*(0,\dots0)=n!$. Hence, as we
reported above, for solving the KPZ equation we need (in our normalisation) 
the value of $ \langle \Phi_0| \mu\rangle/||\mu ||^2$ for arbitrary eigenstates
(even if, at the end of the calculation the choice of normalisation is clearly immaterial). 

In previous works \cite{we-flat,we-flatlong}, we circumvented the problems arising at finite $L$ 
by considering  a wedge initial condition for  (left) half-space problem with partition sum
\be \label{hsdef} 
Z_{\rm hs,w}(x,t) = \int_{-\infty}^{0} dy e^{w y} Z(x,t|y,0)\,,
\ee
with $Z(x,t=0)=\theta(-x) e^{w x}$. 
In the limit $w\to 0^+$ and  $x \to -\infty$ this half-wedge tends to the flat initial condition.
After rather complicated calculations involving this double limit, the following generating function has been 
obtained in Refs. \cite{we-flat,we-flatlong}  (here $\lambda = (\bar c^2 t/4)^{1/3}$):
\be
\label{defg}
\fl g(s) \equiv 1+ \sum_{n} \frac{\overline{Z_{\rm flat}^n(t)}}{n!} (-1)^n e^{- \lambda n s} =
1 + \sum_{\stackrel{N,M \geq 0}{(N,M) \neq (0,0)}} \frac{1}{(2 N)! M!} Z(N,M) ,
\ee
where in the second line we used the fact that the summation over the number of particles 
can be replaced by independent summation over strings $N$ and $M$.
The partition sum at fixed $N,M$  (setting $\cb=1$ for simplicity) is
\bea \label{gener2}
&& \fl Z(N,M) = \sum_{n_1,\dots n_{N}=1}^\infty  \sum_{m_1,\dots m_{M}=1}^\infty
\frac{1}{2^{M}}   (-2)^{\sum_{j=1}^M m_j+ 2 \sum_{p=1}^N n_p} \prod_{j=1}^M \frac{1}{m_j!} 
e^{(m_j^3-m_j)  \frac{\lambda^3}{3}- \lambda m_j s}  \\
&& \fl \times \!\!\! \prod_{1 \leq i<j \leq M} \!\! (-1)^{\min(m_i,m_j)} \frac{|m_i - m_j|}{m_i+m_j} 
 \prod_{p=1}^{N}  \int_{k_p}  \frac{1 }{n_p}
  \prod_{p=1}^N \prod_{q=1}^{n_p} \frac{1}{4 k_p^2 + q^2} 
    \prod_{p=1}^{N}e^{2 (n_p^3-n_p)  \frac{\lambda^3}{3}- 8 n_p k_p^2 \lambda^3 - 2 \lambda n_p s} \nn \\
    && \fl \times 
   \prod_{1\leq p<q\leq N}
\frac{4 (k_p-k_q)^2 +(n_p-n_q)^2}{4 (k_p-k_q)^2 +(n_p+n_q)^2}
\frac{4 (k_p+k_q)^2 +(n_p-n_q)^2}{4 (k_p+k_q)^2 +(n_p+n_q)^2} \nonumber \\
&& \fl \times  (-1)^{ \sum_{p=1}^N n_p}  (2 N -1)!!  \prod_{1 \le p \le N,1 \le j \le M} \frac{4 k_p^2 + (n_p-m_j)^2}{4 k_p^2 + (n_p+m_j)^2}.
 \nn
\end{eqnarray}
where $(2 N -1)!! = (2 N)!/2^N N!$. As already stressed the above generating function was obtained in a rather indirect way and  
the calculations turned out to be complicated and long. 

Alternatively now, using Eq. (\ref{sumf}) we can write, the generating function (\ref{defg}) as a double sum $\sum_n \sum_\mu$
over $n$ and over the parity invariant states at fixed $n$ and in finite size $L$, which, in the limit $L \to +\infty$ leads to
\bea
\fl && ~~~~~ g(s) = 1 +  \sum_{\stackrel{N,M \geq 0}{(N,M) \neq (0,0)}} \frac{1}{M! 2^N N!} \prod_{j=1}^M \prod_{p=1}^{N} \sum_{m_j=1}^\infty  \sum_{n_p=1}^\infty n_p L \int \frac{dk_p}{2 \pi} 
\left.\frac{\langle \Phi_0| \mu\rangle}{||\mu ||^2} \right|_{\mu \equiv \{m_j,n_p,k_p\}} ,
\eea 
where the combinatoric counting factors come from indistinguishability of the string states obtained by (i) permutations of
the $m_j$ (ii) permutations of the $n_p$ (iii) change $k_p \to -k_p$. 
In addition, the summation over centre of mass
momenta of the string behaves as free particle quantisation $\sum_{k_p} \to n_p  L \int \frac{dk_p}{2 \pi}$. 
If we now plug the overlap in Eq. (\ref{ovnorm2}) in the above expression, 
we immediately recover the partition sum in Eq. (\ref{gener2}). This shows the
consistency of the two approaches (finite $L$ and wedge) and amounts to a partial re-derivation of the 
result in Refs. \cite{we-flat,we-flatlong} by the direct method (since for some of the
states we were able to derive the overlap from the de Nardis et al. result, for the
others we used heuristics).
Finally, the later steps 
leading to the probability distribution of $h(x,t)$ and its long time limit are independent of
the regularisation chosen and given in
Refs. \cite{we-flat,we-flatlong} (see also \cite{flatnumerics,d-12,ld-14} for short and long time behaviours).

We mention that the replica approach was used in the literature also to tackle different initial conditions 
and different observables \cite{we,dotsenko,SasamotoStationary,ProlhacNumerics,dg-12,d-13,ld-14} which coincide with 
rigorous solutions of models in the same universality class 
\cite{Johansson2000,spohn2000,ferrari1,reviewCorwin,spohnreview,spohnKPZEdge,corwinDP} when the latter are available.
However, for the flat initial condition, the replica approach is the only successful one up to nowadays,
and rigorous derivation is only presently in progress \cite{quastel-prep}

Let us close this section by a remark on conservation laws. The ``stochastic integrability" of
the KPZ equation stems from the quantum integrability of the Lieb Liniger model.
It thus involves an infinity of local conserved operators in the quantum problem.
In general it is not clear what this implies for the KPZ equation and the equivalent 
directed polymer (DP) problem. Here however,
for the flat initial condition, following the remark by de Nardis et al. \cite{nwbc-13}, we 
learn that the time evolution takes place in the space of parity invariant states
(\ref{parityinv}) 
an exact statement valid for all $t$ and $L$. Without knowing anything else, this 
implies some non trivial properties for correlation functions, e.g.
for $n=3$ (choosing with no loss of generality $x_1<x_2<x_3$): 
\bea
\overline{Z(x_1,t) Z(x_2,t) Z(x_3,t)} = \sum_{1 \leq i<j \leq 3} f_{ij}(x_i-x_j) ,
\eea 
a rather non trivial property, which implies e.g. 
\bea
\overline{\nabla Z(x_1,t) \nabla Z(x_2,t) \nabla Z(x_3,t)} = 0 ,
\eea
for all non coinciding $x_i$ and $t$, which thus act as conserved quantities. 
Similarly all correlations of an odd number of $\nabla Z$ vanish.

\section{Conclusions}
\label{concl}

In this manuscript we investigated some of the deep connections between quantum quench problems 
in the Lieb Liniger model and the KPZ equation with flat initial condition. 
We showed how to generalise the recent conjectured formula \cite{nwbc-13} for the overlap 
between the BEC state and a subset of Bethe eigenstates of the attractive Lieb Liniger model 
(namely the parity invariant, paired string configurations plus a string of zero momentum with even number of particles).
In this way we recovered  some of the results of Refs. \cite{we-flat,we-flatlong} by a completely different method.
We have not yet generalised the fixed size overlap formula to odd number $n$ of particles.
However, also for even $n$ there are other eigenstates of the attractive problem involving zero-momentum 
strings which were unveiled in \cite{we-flat,we-flatlong} for which the fixed size overlap formula is hard to apply. 
However, using the known results for the KPZ flat initial condition problem \cite{we-flat, we-flatlong},
and proceeding by reverse engineering, we conjectured a formula for the 
overlap between the BEC and an arbitrary string state valid at arbitrary fixed and finite $n$ and large $L$.
Using this conjecture we recover the starting formula for the generating function of the probability distribution 
function of the KPZ height, which has been shown \cite{we-flat,we-flatlong} to converge to the 
celebrated Tracy-Widom distribution of the Gaussian Orthogonal Ensemble \cite{TW1994}, as also 
confirmed by recent experiments \cite{exp4,exp5}.

This result represents an alternative and more straightforward derivation of the KPZ generating function 
which circumvents many of the difficult calculations in Ref. \cite{we-flatlong}.
Apart from the technical simplicity, it is also important to have shown that the KPZ equation with flat initial condition can be solved 
without introducing regularisations. 
We still do not know the overlaps for odd number of particles, but, at least in principle,
only the knowledge of the even moments of the partition function (even number of replicas)
should be enough to determine the probability distribution function, although we are not aware 
of any attempt in this direction.

Furthermore, the presented derivation is also a first step towards the solution of the KPZ solution 
in a finite system of length $L$, for which very little is known \cite{bb-00}.
However, in order to achieve this goal, a lot of work is still needed, in particular on how to deal 
with anomalous strings in finite systems and the knowledge of overlap for odd number of particles. 

Apart from the flat initial condition, in the KPZ literature there has been a large interest in the 
so called narrow wedge initial condition \cite{we,dotsenko,spohnKPZEdge,corwinDP}, 
for which, in the replica language, roughly all particles 
are constrained into a single point. It is natural to wonder whether there is a connection between these 
results and quenches from inhomogenous  initial states considered for the Lieb Liniger in Refs. \cite{a-12,ck-12,csc-13}.

Our results also have direct new consequences for the quantum quench problem in the attractive Lieb-Liniger gas. 
Indeed, our conjectured formula for the overlaps of generic string states shows that there are states with coinciding rapidities 
(as e.g. in \ref{AppB}) which have a finite overlap in the infinite volume limit, while their weight in
equilibrium is known to be negligible \cite{cc-07}. 
This is analogous to some other theoretical \cite{fab1} and experimental \cite{bloch1} results
for spin-chains in different quench setups.

There are also a series of physical questions arising from the relation of these two important problems
for which we do not have yet an answer.
Indeed, it has been shown \cite{ce-13,nwbc-13} that the large time behaviour of local observable after a quantum quench 
is described by a Generalised Gibbs Ensemble (GGE) \cite{GGE} in which all the {\it local} integral of motion enter \cite{cef-ii},
as a difference with the standard Gibbs ensemble where only the Hamiltonian matters.
What are the consequences of these conservation laws for the large time behaviour of KPZ equation? 
Conversely, it is well known that the KPZ equation for large time leads to a Tracy-Widom  \cite{TW1994}
probability distribution function. Does the Tracy-Widom distribution enters also in the quantum quench?
Is the GGE somehow connected to Tracy-Widom?

\section*{Acknowledgments}   
PC acknowledge the ERC  for financial  support under  Starting Grant 279391 EDEQS. 
PLD acknowledge the hospitality of the Dipartimento di Fisica dell'Universit\`a di Pisa
where part of this work has been done.


\appendix

\section{Explicit form of the overlaps for small number of particles} 
\label{AppA}

In this Appendix we report the explicit expressions for the overlap of arbitrary parity invariant (un-normalised) Bethe states with 
the (un-normalised) BEC state $\Phi_0(x_1\dots x_n)=1$ for low even and odd number of particles. 

We denote $O_n=\langle \Phi_0|\Psi_\mu(x_1\dots x_n)\rangle$. We 
insert the parity invariance condition (\ref{parityinv}) in the expression of the 
Bethe states (\ref{def1}), and in the Bethe equations (\ref{BE}) (which reduce
to a smaller set of equations). We then carry an 
explicit calculation of the overlap integral which is simplified using the (reduced)
Bethe equations, using mathematica. We find:
\bea
\fl && O_2  = \frac{2 c L}{\lambda _1^2} \\
\fl && O_3 = \frac{6 c L^2 \left(c^2+\lambda _1^2\right)}{\lambda _1^4}+\frac{36 c^2 L}{\lambda _1^4} \\
\fl && O_4 = \frac{24 c^2 L^2  \left(c^2+\left(\lambda _1-\lambda _2\right){}^2\right)
   \left(c^2+\left(\lambda _1+\lambda _2\right){}^2\right)}{\lambda
   _1^2 \lambda _2^2 \left(\lambda _1^2-\lambda _2^2\right){}^2}+\frac{192
   c^3 L \left(c^2+\lambda _1^2+\lambda _2^2\right)}{\lambda _1^2 \lambda
   _2^2 \left(\lambda _1^2-\lambda _2^2\right){}^2} \label{O4} \\
   \fl && O_5 = \frac{120 c^2 L^3 \left(c^2+\lambda _1^2\right) \left(c^2+\lambda
   _2^2\right)\left(c^2+\left(\lambda _1-\lambda _2\right){}^2\right)
   \left(c^2+\left(\lambda _1+\lambda _2\right){}^2\right) }{\lambda
   _1^4 \lambda _2^4 \left(\lambda _1^2-\lambda _2^2\right){}^2}\\
\fl   && +\frac{240 c^3
   L^2 \left(10 c^6+23 c^4 \left(\lambda _1^2+\lambda _2^2\right)+4 c^2
   \left(4 \lambda _1^4+3 \lambda _2^2 \lambda _1^2+4 \lambda _2^4\right)+3
   \lambda _1^6+3 \lambda _2^6+\lambda _1^2 \lambda _2^4+\lambda _1^4
   \lambda _2^2\right)}{\lambda _1^4 \lambda _2^4 \left(\lambda
   _1^2-\lambda _2^2\right){}^2} \nn \\
\fl      && +\frac{2400
   c^4 L \left(5 c^4+8 c^2 \left(\lambda _1^2+\lambda _2^2\right)+3 \lambda
   _1^4+3 \lambda _2^4+2 \lambda _1^2 \lambda _2^2\right)}{\lambda _1^4
   \lambda _2^4 \left(\lambda _1^2-\lambda _2^2\right){}^2} \nn
\eea 
We have also exact results for $n=6$ and $n=7$, but their expression are too long to be reported here.
For $n=2,4,6$ these overlaps coincide for arbitrary complex value of $\lambda$'s 
with the conjecture in Eq. (\ref{overlap2}).
Note that $O_2$ was given in Eq. (50) of Ref. \cite{we-flatlong}.

\section{An anomalous string}
\label{AppB}

In this Appendix we consider the simplest state with coinciding rapidities in the infinite volume limit, namely the state 
with $M=2$, $m_1=1$ and $m_2=3$ (and $N=0$). 
This state, in the infinite volume limit, is composed by two zero momentum strings
with rapidities' set $(ic,0,-ic)$ and $(0)$.  
Such a state is naively not allowed, since two rapidities coincide, but we can construct is as the limit 
of a finite volume parity invariant state with rapidities 
$\lambda_1= i c + w^2 \delta_1$ and $\lambda_2 = w \delta_2$, and clearly $\l_3=-\l_1$, $\l_4=-\l_2$. 
In this appendix we denote $w = e^{-\bar c L/2}$ (not to be confused with $w$ in the text)
which represents the string deviation, and $\delta_{1,2}$ are amplitude that 
will be fixed by solving finite volume Bethe equations (in other cases, these string deviations
do not matter because they are small, but now they should be taken into account to avoid coinciding rapidities
and hence divergences in norms and overlaps).
Using the exact formula for the norm and the overlap in finite size for $4$ particles (e.g. Eqs. (\ref{Gaudin}) and (\ref{O4})) 
we have at the first non vanishing order in $w$   
\bea
 ||\mu||^2 &=& 4!^2 ( - \frac{L}{\bar c^3} + 2 i \frac{\delta_1 L}{c^2 \delta_2^2} + \frac{L^2}{2 c^2})  + O(w), \label{B1}\\
 \langle \Phi_0|\mu \rangle &=&  -192 ( - \frac{L}{\bar c^3} + 2 i \frac{\delta_1 L}{c^2 \delta_2^2} + \frac{L^2}{2 c^2})  + O(w) \label{B2}. 
\eea 
Taking the ratio $ \langle \Phi_0|\mu \rangle/ ||\mu||^2$ we obtain 
$-1/3$ (independently of $\delta_{1,2}$, $L$, and $\cb$) 
which is the correct result in agreement with our Eq. (\ref{ovnorm2}).
Thus we do not need to use the Bethe equations to calculate this ratio. 
However each of the two terms must have a defined infinite volume limit going like $L^2$ times
a $\delta$ independent factor. 
In order to fix these coefficients, we need to solve the Bethe equations.
After some simple algebra we obtain the solution
\bea
\delta_1 = - 6 i c^2 L, \qquad \delta_2^2 = 12 c^2. 
\eea 
Plugging this in the above result for the norm we have at the leading order in $L$
\bea
||\mu||^2 = 36 \times 4! L^2/c^2 
\eea 
which is also the result one obtains from the norm formula (\ref{norm3}) setting $M=2$, $N=0$
$m_1=1, m_2=3$. 
The result for the overlap is then straightforwardly  $-1/3$ times the norm squared.

Notice how plugging in the solution to Bethe equation in Eqs. (\ref{B1}) and (\ref{B2}) the second term 
proportional to $\delta_1 L$ becomes $\propto L^2$, showing that the limit $L\to \infty$ and 
$w\to 0$ do not commute and that they should be handled with care, which is the main problem with these 
anomalous states.

\end{document}